\documentclass[aps,pre,reprint,nofootinbib]{revtex4-1} 

\usepackage{amsthm}
\usepackage{amssymb}   
\usepackage{mathtools} 
\usepackage{hyperref}
\hypersetup{colorlinks=true,linkcolor=blue,urlcolor=blue,citecolor=blue}
\usepackage{accents}
\usepackage{tensor}
\usepackage[cal=boondox]{mathalfa}
\usepackage{lipsum}

\usepackage[T1]{fontenc}

\newcommand{\rmi}{{\rm i }}
\usepackage{mathtools}
\DeclarePairedDelimiter\bra{\langle}{\rvert}
\DeclarePairedDelimiter\ket{\lvert}{\rangle}
\DeclarePairedDelimiterX\braket[2]{\langle}{\rangle}{#1 \delimsize\vert #2}

\begin{document}

\title{Classical analog of the quantum metric tensor}

\author{Diego Gonzalez}
\email{diego.gonzalez@correo.nucleares.unam.mx}

\author{Daniel Guti\'errez-Ruiz}
\email{daniel.gutierrez@correo.nucleares.unam.mx}

\author{J. David Vergara}
\email{vergara@nucleares.unam.mx}

\affiliation{Departamento de F\'isica de Altas Energ\'ias, Instituto de Ciencias Nucleares, Universidad Nacional Aut\'onoma de M\'exico, Apartado Postal 70-543, Ciudad de M\'exico, 04510, M\'exico}

\date{\today}

\begin{abstract}
We present a classical analog of the quantum metric tensor, which is defined for classical integrable systems that undergo an adiabatic evolution governed by slowly varying parameters. This classical metric measures the distance, on the parameter space, between two infinitesimally different points in phase space, whereas the quantum metric tensor measures the distance between two infinitesimally different quantum states. We discuss the properties of this metric and calculate its components, exactly in the cases of the generalized harmonic oscillator, the generalized harmonic oscillator with a linear term, and perturbatively for the quartic anharmonic oscillator. Finally, we propose alternative expressions for the quantum metric tensor and Berry's connection in terms of quantum operators.
\end{abstract}

\maketitle

\section{Introduction}

Two fundamental structures for understanding the geometrical aspects of quantum states are the quantum metric tensor formulated by Provost and Vallee~\cite{Provost1980,Wootters1981} and the geometric phases, in particular, the phase discovered by Berry~\cite{Berry45}. The quantum metric tensor is defined in the parameter space and measures the distance between two states corresponding to infinitesimally different parameters. Remarkably, the singularities of this metric are associated with quantum phase transitions exhibited by the corresponding system~\cite{Zanardi2007,SHI-JIAN2010}.  Further, the geodesics induced by this metric can also indicate the presence of quantum phase transitions~\cite{Kumar2012,Kumar2014}. In general, the quantum metric tensor played an essential role in diverse physical phenomena (see Ref.~\cite{Ozawa2018} and references therein). Berry's phase is the extra phase acquired by the wave function when the system undergoes an adiabatic excursion along a closed path in the parameter space and can be understood as an integral of a curvature~\cite{Simon1983}, the so-called Berry curvature. This phase was analyzed in various contexts~\cite{Wilczek1984,Zhang2005,Mikko2008,Xiao2010},  and, interestingly, it is also connected with quantum phase transitions~\cite{Zhu2006}. These approaches to quantum phase transitions based on the metric and the Berry phase can be unified in terms of the critical singular behavior of the quantum geometric tensor~\cite{VenutiZanardi2007,SHI-LIANG2008}, whose real part gives the quantum metric tensor whereas the imaginary part gives the Berry curvature. 

On the other hand, Berry's phase possesses a classical counterpart known as Hannay's angle~\cite{Hannay1985}. For classical integrable systems, it is an extra angle shift picked up by the angle variables of the system when the parameters undergo a closed adiabatic excursion in the parameter space. This classical angle was investigated in a variety of systems~\cite{Khein1993,BerryMorgan1996,Golin1989,Chattopadhyay2018}, and the semiclassical relation between it and Berry's phase was established in Ref.~\cite{Berry1985} and has been verified in many systems~\cite{Berry1985,Datta1989,Biswas1990,Brihaye1993}.

In the light of this and given the close relationship between the quantum metric tensor and Berry's curvature, a natural question arises: What about the classical analog of the quantum metric tensor? It is well known that, in the context of thermodynamic systems, Weinhold~\cite{Weinhold1975} and later Ruppeiner~\cite{Ruppeiner1979} proposed classical metrics in the parameter space which are defined as the Hessian of a thermodynamic potential. For Weinhold's metric, the potential is the internal energy, whereas for Ruppeiner's metric the potential is the entropy. In spite of the existence of these classical metrics, there is so far no evidence for that in the context of classical mechanical systems.

In this paper, we present a meaningful metric tensor for classical integrable systems, which is defined in the parameter space and is the classical analog of the quantum metric tensor. These metrics are analogous in the sense that both yield the same parameter structure, modulo the use of the Bohr-Sommerfeld quantization rule for action variables. It means that we can extract out the same (or almost the same) ``relevant'' information from either of these metrics. This important feature will be exhibited by the three examples that we have considered: the generalized harmonic oscillator, the generalized harmonic oscillator with a linear term, and the quartic anharmonic oscillator. Another important property of this classical metric, which is shared with Hannay's angle, is that it is gauge invariant in the parameter space in that it does not depend on the choice of the point of origin from which we measure the angle variables. The fundamental building blocks from which the classical metric is constructed are certain functions that generate displacements in the parameter space. By promoting these classical functions to quantum operators, we also find alternative expressions for the quantum metric tensor and Berry's connection.

The paper is organized as follows.  In Sec.~\ref{sec:QIM} we briefly review some basics about the quantum metric tensor. In Sec.~\ref{sec:classical} we define  the notion of distance on the parameter space between points in phase space and derive the classical analog of the quantum metric tensor. In Sec.~\ref{Examples} we compute and compare this classical metric and the quantum metric tensor for the considered systems. Section \ref{sec:alternative} presents alternative expressions  for the quantum metric tensor and Berry's connection. Finally, Sec. \ref{sec:Conclusions}  is devoted to conclusions and directions for future research.

\section{Quantum metric tensor}\label{sec:QIM}

In this section, we shortly review  the definition of the quantum metric tensor. We start by considering a quantum theory defined by a set of phase space operators $\hat{q}=\{\hat{q}^a\}$ and $\hat{p}=\{\hat{p}_a\}$ ($a,b,\dots\! =1,\dots,n$) together with a  Hamiltonian operator $\hat{H}(\hat q, \hat p ; x)$,  that depends of this set and also smoothly depends on a set of $N\geq2$ external parameters $x=\{x^i\}$ ($i,j,\dots\!=1,\dots,N$) that are regarded as slowly varying functions of the time~$t$ (adiabatic parameters) and parametrize some $N$-dimensional parameter manifold  $\mathcal{M}$. Assuming that $\hat{H}[x(t)]$ has at least one eigenvector $\ket{n[x(t)]}$ with nondegenerate eigenvalue $E_n[x(t)]$, the adiabatic theorem states that if the system is initially prepared in $\ket{n[x(0)]}$, then during the quantum adiabatic evolution it will remain in the same state $\ket{n[x(t)]}$. This fact means that, under the small change of points $x \rightarrow x'=x+\delta x$ in $\mathcal{M}$, the state $\ket{n(x)}$ will become $\ket{n(x')}$. In consequence, the distance between the states $\ket{n(x)}$ and $\ket{n(x')}$ is defined by 
\begin{equation}\label{QIMdistance}
dl^2\equiv 1 - | \braket{n(x)}{n(x')} |^2,
\end{equation}
where $f=| \braket{n(x)}{n(x')}|$ is the \textit{fidelity} and measures the similarity between states. After expanding $\ket{n(x')}$ into a second-order Taylor series, Eq.~(\ref{QIMdistance}) can be expressed as $dl^2\simeq g^{(n)}_{ij} \delta x^i \delta x^j$ where
\begin{eqnarray}\label{QIM}
g^{(n)}_{ij}(x) \equiv {\rm Re} \left(\braket{\partial_i n}{\partial_j n} - \braket{\partial_i n}{n}\braket{ n}{\partial_j n}\right), 
\end{eqnarray}
is the \textit{(abelian) quantum metric tensor}~\cite{Provost1980}. 
An alternative expression for this metric derived from the Lagrangian formalism is given in Ref.~\cite{Alvarez-Jimenez2017}. Throughout this paper, we adopt the convention that repeated indices $i,j,\dots,$ are summed from $1$ to $N$, and $\partial_i := \partial/ \partial x^i$.

For the purposes of this paper, it is convenient to cast Eq.~(\ref{QIM}) in terms of operators. Let $\hat{P}_i$ be Hermitian operators and consider that $\hat{P}_i\delta x^i$ is the generator of the displacement $\ket{n(x)}\rightarrow\ket{n(x')}$. Thus, the translated state is 
\begin{equation}\label{QIMOperator}
\ket{n(x')}=\exp\left(-\frac{\rmi}{\hbar} \delta x^i\hat{P}_i\right)\ket{n(x)}.
\end{equation}
From this equation, by considering a Taylor expansion, we have 
\begin{equation}\label{QIMOperator2}
\rmi \hbar \ket{\partial_i n(x)}=\hat{P}_i \ket{n(x)},
\end{equation}
which substituted into Eq.~(\ref{QIM}) leads to~\cite{Provost1980} 
\begin{equation}\label{QIM2}
g^{(n)}_{ij}(x) = \frac{1}{\hbar^2} {\rm Re} \left( \langle \hat{P}_i \hat{P}_j \rangle_n -\langle \hat{P}_i \rangle_n \langle \hat{P}_j\rangle_n \right), 
\end{equation}
where $\langle \hat{X} \rangle_n\equiv\bra{n}\hat{X}\ket{n}$ is the expectation value of $\hat{X}$ with respect to the state $\ket{n}$. It should be noted that because of the Hermiticity of $\hat{P}_i$, the right-hand side (r.h.s) of Eq.~(\ref{QIM2}) is symmetric. Furthermore, the line element $dl^2 = g^{(n)}_{ij} \delta x^i \delta x^j$ now reads 
\begin{equation}\label{Qdistance}
dl^2=\frac{1}{\hbar^2 }\langle \Delta \hat{P}^2 \rangle_n, \qquad (\Delta \hat{P}=\Delta \hat{P}_i \delta x^i),
\end{equation}
where $\Delta \hat{P}_i:=\hat{P}_i-\langle \hat{P}_i \rangle_n$. Then, using operators, the distance $dl^2$ can be seen as the variance of the generator $\hat{P}_i\delta x^i$. This last
remark will be the key point to obtain the classical counterpart of the quantum metric in the next section.

\section{Classical analog of the quantum metric tensor}\label{sec:classical}

We now turn to the classical setting. Let us consider a classical integrable system with $n$ degrees of freedom described by the time-dependent Hamiltonian $H[q,p;x(t)]$, where $q=\{q^a\}$ and $p=\{p_a\}$ are the canonical coordinates and  momenta, and $x=\{x^i\}\in\mathcal{M}$ is the set of slow time-dependent parameters.

Since the system is integrable (for all values of $x \in \mathcal{M}$), we can introduce the action-angle variables, $I=\{I_a\}$ and $\varphi=\{\varphi^a\}$, which satisfy Hamilton's equations of motion with the new Hamiltonian 
\begin{equation}\label{classical:K}
K(\varphi,I;x)=H(I;x) - G_i(\varphi,I;x) \dot{x}^i,
\end{equation}
where $H(I;x) \equiv H[q(\varphi,I;x),p(\varphi,I;x);x]$ depends only on the action variables and the parameters, and $G_i(\varphi,I;x):= G_i[q(\varphi,I;x),I;x]$ with
\begin{equation}\label{classical:G}
G_i(q,I;x) := - ( \partial_i S^{(\alpha)} )_{q,I},
\end{equation}
where $S^{(\alpha)}(q,I;x)$ is the generating function of the canonical transformation $(q,p) \rightarrow (\varphi,I)$. Also, $\dot{x}^i := dx^i/dt$ and $\alpha$ label different branches of the  multivalued function $S^{(\alpha)}(q,I;x)$.  We recall that the second term in the r.h.s of Eq.~(\ref{classical:K}) comes from $( \partial S^{(\alpha)}/ \partial t)_{q,I}=(\partial S^{(\alpha)}/\partial x^{i})_{q,I} \dot{x}^i$, which is a consequence of the fact that $H[q,p,x(t)]$ (and hence $S^{(\alpha)}[q,I;x(t)]$ also) depends explicitly on time through the parameters $x$. The explicit form of $G_i$ in terms of the action-angle variables is
\begin{equation}\label{classical:G2}
G_i(\varphi,I;x) = p_a ( \partial_i q^a)_{\varphi,I} - ( \partial_i S )_{\varphi,I},
\end{equation} 
where $p_a=p_a(\varphi,I;x)$, $q^a=q^a(\varphi,I;x)$ and we defined the single-valued function $S(\varphi,I;x):=S^{(\alpha)}[q(\varphi,I;x),I;x]$ with $0\leq \varphi <  2\pi$. We use the notation that repeated indices $a,b,\dots,$ are summed from $1$ to $n$.

As  our first step towards the classical counterpart of Eq.~(\ref{QIM2}), we find that under the action of an infinitesimal displacement of the parameters $x \rightarrow x'=x+\delta x$ in $\mathcal{M}$, the function $G_i \delta x^i$ is the generator of the infinitesimal canonical transformation 
\begin{equation}\label{classical:inf}
[q(x),p(x)] \rightarrow [q(x)+\bar{\delta}q ,p(x)+\bar{\delta}p], 
\end{equation}
where
\begin{subequations}
\begin{eqnarray}
\bar{\delta}q^a:=q^a(x')-q^a(x)=( \partial_i q^a )_{\varphi,I} \delta x^i, \label{classical:deltaq}\\
\bar{\delta}p_a:=p_a(x')-p_a(x)=( \partial_i p_a )_{\varphi,I} \delta x^i.\label{classical:deltap}
\end{eqnarray}
\end{subequations}
Notice that another form of Eq.~(\ref{classical:deltaq}) is $\bar{\delta}q^a=\delta q^a-\tilde{\delta} q^a$, where $\delta q^a := q'^a(x')-q^a(x)$ is the total variation and $\tilde{\delta} q^a := q'^a(x)-q^a(x)$ is the variation with ``frozen'' parameters. A similar expression follows for $\bar{\delta}p_a$. 

To prove the above statement it is sufficient to show that $G_i$ satisfy
 \begin{subequations}
	\begin{eqnarray} 
	( \partial_i q^a )_{\varphi,I}&=& \{ q^a , G_i \}_{q,p} =\frac{\partial G_{i}}{\partial p_{a}}, \label{classical:Gq}\\
	\left( \partial_i p_a \right)_{\varphi,I}&=& \{ p_a , G_i \}_{q,p}=-\frac{\partial G_{i}}{\partial q^{a}}, \label{classical:Gp}
	\end{eqnarray} 
\end{subequations}
which are the equations of the infinitesimal canonical transformation (\ref{classical:inf})~\cite{KOLODRUBETZ20171}. Here $\{ \cdot , \cdot \}$ denotes the Poisson bracket. To do this, we first take the partial derivative with respect to $x^i$, holding $(\varphi,I)$ fixed, of the familiar relation  $p_a \tilde{d} q^a - I_a \tilde{d} \varphi^a = \tilde{d} F$, where $F=S^{(\alpha)}(q,I;x)-I_a \varphi^a$ and $\tilde{d}$ is the fixed-time differential (or equivalently with fixed parameters $x$). From this we obtain
\begin{equation}\label{classical:dG1}
\left( \partial_i p_a\right)_{\varphi,I} \tilde{d} q^a + p_a \tilde{d} \left( \partial_i q^a \right)_{\varphi,I} = \tilde{d} \left( \partial_i S \right)_{\varphi,I},
\end{equation}
where we used $(\partial_i \tilde{d} f)_{\varphi,I} =\tilde{d}(\partial_i f)_{\varphi,I} $. Next, combining Eq.~(\ref{classical:dG1}) with the differential of Eq.~(\ref{classical:G2}) at fixed $x$, namely
\begin{equation}
\tilde{d}G_i=\tilde{d}p_a ( \partial_i q^a)_{\varphi,I} +p_a \tilde{d}( \partial_i q^a)_{\varphi,I}- \tilde{d}( \partial_i S )_{\varphi,I},
\end{equation}
we have
\begin{equation}\label{classical:G3}
\tilde{d}G_i=- \left( \partial_i p_a \right)_{\varphi,I}  \tilde{d} q^a + \left( \partial_i q^a \right)_{\varphi,I} \tilde{d} p_a. 
\end{equation}
Then, taking $G_i$ as a function of $(q,p)$, it follows that
\begin{equation}\label{classical:G4}
\tilde{d}G_i=\frac{\partial G_i}{\partial q^a} \tilde{d} q^a + \frac{\partial G_i}{\partial p_a} \tilde{d} p_a.
\end{equation}
Equating the coefficients of  $\tilde{d} q^a$ and $\tilde{d} p_a$ on the r.h.s of Eqs.~(\ref{classical:G3}) and (\ref{classical:G4}), we read off Eqs.~(\ref{classical:Gq}) and (\ref{classical:Gp}), which completes the proof. 

Given the fact that $G_i \delta x^i$ generates an infinitesimal displacement in $x$ of points in phase space, and in complete analogy with the quantum case [see Eq.~(\ref{Qdistance})], we can naturally define the distance between the points  $[q(x),p(x)]$ and $[q(x)+\bar{\delta}q ,p(x)+\bar{\delta}p]$ as
\begin{equation}\label{gclas:distance}
ds^2:=\left<\Delta G^2\right>\qquad (\Delta G=\Delta G_i \delta x^i),
\end{equation}
where $\Delta G_i :=G_i - \left<G_i\right>$ and
\begin{equation}
\left< f(\varphi,I;x)  \right>=\frac{1}{(2 \pi)^{n}}\oint d\varphi f(\varphi,I;x),
\end{equation}
with $\oint d\varphi =\prod_{a=1}^{n} \int_{0}^{2 \pi}d\varphi^a$, is the average of $f(\varphi,I;x)$ over the (fast) angle variables. Defined in this way, the classical distance $ds^2$ is nothing more than the variance of the generator $G_i \delta x^i$. Clearly, if the parameters $x$ are frozen, then $G_i\delta x^i=0$, and hence $ds^2$ also vanishes, as expected.

Notice that $ds^2$ depends only on the action variables~$I$ and the parameters~$x$. In this regard, it is important  to emphasize that, according to the classical adiabatic theorem~\cite{Arnold2006}, while the parameters vary slowly with time, the action variables are adiabatic invariants\footnote{Nevertheless, for Hamiltonian systems with $n\geq2$ there may exist conditions for which the adiabatic approximation is not optimal~\cite{Arnold2006}.} $\dot{I}_a\approx0$. That is, during the adiabatic evolution from $[q(x),p(x)]$ to $[q(x)+\bar{\delta}q ,p(x)+\bar{\delta}p]$ the action variables $I$ remain constant. This effect is similar to the quantum case where the quantum number $n$ remains constant  as the parameters vary. On the other hand, note also that in this scenario, the average $\left< \cdot \right>$ in Eq.~(\ref{gclas:distance}) is the classical counterpart of the quantum average $\langle \cdot \rangle_n$ in Eq.~(\ref{Qdistance}).

By expanding Eq.~(\ref{gclas:distance}), we find that the distance $ds^2=g_{ij} \delta x^i \delta x^j$ induces the metric 
\begin{equation}\label{gclas:metric}
g_{ij}(I;x):=\left< G_i G_j\right> - \left<G_i\right> \left<G_j\right>,  
\end{equation}
where $G_i=G_i(\varphi,I;x)$ is given by Eq.~(\ref{classical:G2}). The metric $g_{ij}(I;x)$ corresponds to the \textit{classical analog of the quantum metric tensor}~(\ref{QIM}) [or Eq.~(\ref{QIM2})], and provides a measure of the distance between the nearby points $[q(x),p(x)]$ and $[q(x)+\bar{\delta}q ,p(x)+\bar{\delta}p]$ on the parameter manifold  $\mathcal{M}$. It should be pointed out that, in contrast to the quantum metric tensor, the classical metric (\ref{gclas:metric}) is restricted to the case where classical motion is integrable. This restriction is to be expected since it is the same as that found in Hannay's angle~\cite{Hannay1985}, which also involves the action variables and is the classical counterpart of Berry's phase~\cite{Berry1985}.

We now proceed to check some properties of $g_{ij}(I;x)$. Let us first show that, under a coordinate transformation, $g_{ij}(I;x)$ transforms as a tensor. By considering a coordinate change $y = y(x)$ and using Eq.~(\ref{classical:G2}), it follows that the transformation law for $G_i$ is 
\begin{equation}
G'_{i}(\varphi,I;y)=\frac{\partial x^j}{\partial y^i} G_{j}[\varphi,I;x(y)].
\end{equation}
This result, together with Eq.~(\ref{gclas:metric}), leads to the expected transformation law for the metric
\begin{equation}
g'_{ij}(I;y)=\frac{\partial x^k}{\partial y^i} \frac{\partial x^l}{\partial y^j} g_{kl}[I;x(y)].
\end{equation}
We now prove that $g_{ij}(I;x)$ is positive semidefinite.  This is straightforward and follows from the fact that $ds^2=\left<\Delta G^2\right>\geq0$ since the variance is nonnegative.  In this light, it is interesting to note that the quantum metric tensor (\ref{QIM}) is also positive semidefinite~\cite{chruscinski2012geometric,amari2016information}.

Analogously as the quantum metric $g^{(n)}_{ij}(x)$ is independent of the gauge transformation\footnote{In Ref.~\cite{Alvarez-Jimenez2016} is shown, however, that under a more general  gauge transformation, the quantum metric tensor depends on the gauge.} $\ket{n'(x)} = \exp[i \alpha_n(x)] \ket{n(x)}$ where $\alpha_n(x)$ is an arbitrary real function of $x$, the classical metric $g_{ij}(I;x)$ is invariant under the (gauge) canonical transformation
\begin{equation}\label{gclas:gauge}
\varphi'^a = \varphi^a + \frac{\partial \lambda(I;x)}{\partial I_a}, \qquad I'_a=I_a,
\end{equation}
which is generated by the function $F_2=\varphi^a I'_a + \lambda(I';x)$ where $\lambda(I';x)$ is an arbitrary function of $I'$ and $x$. The proof of this statement is as follows. The Hamiltonian for the new action-angle variables $(\varphi',I')$ is
\begin{equation}
K'(\varphi',I';x)=H(I';x) - G'_i(\varphi',I';x) \dot{x}^i,
\end{equation}
where $H(I';x)=H(I;x)$ with $I'=I$, and 
\begin{equation}\label{gclas:Gprima}
G'_i(\varphi',I';x)=G_i(\varphi',I';x)-[\partial_i \lambda(I';x)]_{I'}, 
\end{equation}
where $G_i(\varphi',I';x):=G_i[\varphi(\varphi',I';x),I';x]$ are the functions $G_i$ for $(\varphi,I)$ expressed in terms of the variables $(\varphi',I')$. Since $G'_i(\varphi',I';x)$ satisfy Eqs.~(\ref{classical:Gq}) and (\ref{classical:Gp}) with $(\varphi',I')$ instead of $(\varphi,I)$, it follows that $G'_i \delta x^i$ generates a canonical transformation of the same type as Eq.~(\ref{classical:inf}) with $\bar{\delta}'q^a=( \partial_i q^a )_{\varphi',I'} \delta x^i$ and $\bar{\delta}'p_a=( \partial_i p_a )_{\varphi',I'} \delta x^i$ instead of Eqs.~(\ref{classical:deltaq}) and (\ref{classical:deltap}), respectively. With this in mind, we can apply Eq.~(\ref{gclas:metric}), and write the classical metric associated with the variables $(\varphi',I')$ as
\begin{equation}\label{gclas:metric2}
g'_{ij}(I';x)=\left< G'_i G'_j\right>' - \left<G'_i\right>' \left<G'_j\right>',  
\end{equation}
where $G'_i=G'_i(\varphi',I';x)$ and $\left< \cdot \right>'$ stands for the average over the angle variables $\varphi'$. 

By using Eq.~(\ref{gclas:Gprima}), the average $\left<G'_i\right>'$ gives
\begin{eqnarray}\label{gclas:promGprima}
&&\left<G'_i(\varphi',I';x)\right>' \!=\!\frac{1}{(2 \pi)^{n}}\oint d\varphi' G_i(\varphi',I';x)\!-\![\partial_i \lambda(I';x)]_{I'} ,\nonumber\\
&&=\frac{1}{(2 \pi)^{n}} \int_{-b_n}^{2 \pi-b_n}\dots \int_{-b_1}^{2 \pi-b_1} d\varphi^1 \dots d\varphi^n G_i(\varphi,I;x)  \nonumber \\
&&\ \ \ -[\partial_i \lambda(I;x)]_{I}, \nonumber\\
&&= \left<G_i(\varphi,I;x)\right> -[\partial_i \lambda(I;x)]_{I}, 
\end{eqnarray}
where in the second line we made the change of variables from $\varphi'$ to $\varphi$ and defined $b_a:=\partial \lambda(I;x) / \partial I_a$, whereas in the last line we used the fact that $p$, $(\partial_i q)_{\varphi,I}$, and $(\partial_i S)_{\varphi,I}$ are periodic functions of each angle variable $\varphi^a$ with period $2\pi$, which by virtue of Eq.~(\ref{classical:G2}) implies that $G_i(\varphi,I;x)$ are also periodic functions of each $\varphi^a$. The periodicity of $(\partial_i S)_{\varphi,I}$ is easily seen by writing  $S(q,I;x)=\sum_{a=1}^{n}S_{a}(q_a,I;x)$ and recalling that each $S_{a}(\varphi,I;x)\equiv S_{a}[q_a(\varphi,I;x),I;x]$ satisfies $S_a(\varphi+2\pi,I;x)-S_a(\varphi,I;x)=2\pi I_a$. In the same fashion, the average $\left<G'_iG'_j\right>'$ leads to
\begin{eqnarray}\label{gclas:promGGprima}
&&\left<G'_i(\varphi',I';x)G'_j(\varphi',I';x)\right>'=\left<G_i(\varphi,I;x)G_j(\varphi,I;x)\right> \nonumber \\
&& - \left<G_i(\varphi,I;x)\right>  [\partial_j \lambda(I;x)]_{I} - \left<G_j(\varphi,I;x)\right>  [\partial_i \lambda(I;x)]_{I} \nonumber \\
&& + [\partial_i \lambda(I;x)]_{I} [\partial_j \lambda(I;x)]_{I}. 
\end{eqnarray}
It remains to  substitute Eqs.~(\ref{gclas:promGprima}) and (\ref{gclas:promGGprima}) into Eq.~(\ref{gclas:metric2}). By doing so, all the terms involving the derivatives of $\lambda(I;x)$  cancel among themselves and thus the metric $g'_{ij}(I';x)$ becomes
\begin{equation}\label{gclas:metricgauge}
g'_{ij}(I';x)=\left< G_i G_j\right> - \left<G_i\right> \left<G_j\right>= g_{ij}(I;x),
\end{equation}
which is the desired result. 

Therefore, although the angle variables are not unique but only defined up to the canonical transformation (\ref{gclas:gauge}), the metric $g_{ij}(I;x)$ is unique and independent of this (gauge) transformation, as expected for a metric tensor on~$\mathcal{M}$. It is interesting to note from Eq.~(\ref{gclas:promGGprima}) that term $\left< G_i G_j\right>$ is not invariant under the transformation~(\ref{gclas:gauge}), and hence it cannot be used alone to define a metric on~$\mathcal{M}$. As shown above, it must be combined with $\left<G_i\right> \left<G_j\right>$, in the precise form given by Eq.~(\ref{gclas:metric}), to produce a gauge invariant metric. This is the essence of the nontrivial gauge invariance of $g_{ij}(I;x)$ under Eq.~(\ref{gclas:gauge}); it arises as a consequence of the particular combination of both $\left< G_i G_j\right>$ and $\left<G_i\right> \left<G_j\right>$. Reinforcing the analogy made between the classical metric $g_{ij}(I;x)$ and the quantum metric tensor  $g^{(n)}_{ij}(x)$, since the latter being a combination of $\braket{\partial_i n}{\partial_j n}$ and  $\braket{\partial_i n}{n}\braket{ n}{\partial_j n}$ is gauge invariant, but the term $\braket{\partial_i n}{\partial_j n}$ alone is not~\cite{Provost1980}.

To end this section, let us add some comments on the significance of $G_i$. Notice that Eq.~(\ref{gclas:Gprima}) reveals that under Eq.~(\ref{gclas:gauge}) the functions $G_i$ transform as an abelian gauge potential, which is not surprising since these functions are the generators of translations in~$\mathcal{M}$~\cite{KOLODRUBETZ20171}. The average of $G_i$ can be identified as the components of the connection 1-form on $\mathcal{M}$ associated with Hannay's angle, namely $A(I;x)=A_i dx^i$ with\footnote{In the literature, however, it is often found that the term $\left<( \partial_i S )_{\varphi,I}\right>$ is dropped from Eq.~(\ref{gclas:connection}) since it does not contribute to Hannay's angle~\cite{Gozzi1987,chruscinski2012geometric}.}
\begin{equation}\label{gclas:connection}
A_i(I;x):=\left<G_i(\varphi,I;x)\right>=\left<p_a ( \partial_i q^a)_{\varphi,I}\right> - \left<( \partial_i S )_{\varphi,I}\right>. 
\end{equation}
This means, according to Eq.~(\ref{gclas:promGprima}), that under the transformation (\ref{gclas:gauge}) the components (\ref{gclas:connection}) transform as those of an abelian gauge potential~\cite{Littlejohn1988}
\begin{equation}
A'_i(I';x)=A_i(I;x)-[\partial_i \lambda(I;x)]_{I}.
\end{equation}
Besides, it follows from Eq.~(\ref{gclas:connection}) that the curvature 2-form $F(I;x)=dA(I;x)$ of this connection can be written as $F(I;x)=(1/2) F_{ij}dx^i\wedge dx^j$ with
\begin{equation}\label{gclas:Curv}
F_{ij}(I;x)=\left< \left( \partial_i p_a \right)_{\varphi,I} \left( \partial_j q^a \right)_{\varphi,I} - \left( \partial_j p_a \right)_{\varphi,I} \left( \partial_i q^a \right)_{\varphi,I}\right>.
\end{equation}
Upon using Eqs. (\ref{classical:Gq}) and (\ref{classical:Gp}), we find that these components take the form
\begin{equation}\label{gclas:curvature}
F_{ij}(I;x)=-\left< \{ G_i , G_j \}_{q,p} \right>.
\end{equation}
In this way the functions $G_i$ can be regarded as the fundamental building blocks that underlie the classical metric~(\ref{gclas:metric}) and Hannay's curvature~(\ref{gclas:curvature}). 

\section{Illustrative  examples}\label{Examples}

In this section, we set out some examples of classical integrable systems to illustrate the appearance of the metric~(\ref{gclas:metric}). At the same time, we compare the results of this  classical metric with those found by using the quantum metric tensor~(\ref{QIM}) associated with the quantum counterpart of each system. We shall see that these results corroborate that the metric~(\ref{gclas:metric}) is the classical analog of~(\ref{QIM}) [or equivalently Eq.~(\ref{QIM2})]. 

\subsection{Generalized harmonic oscillator}

As our first example, let us take the generalized harmonic oscillator, whose classical Hamiltonian is given by
\begin{equation}\label{gho:classicalH}
H=\frac{1}{2}\left(X q^2+2Yqp+Zp^2\right),
\end{equation}
where $x=\{x^{i}\}=(X,Y,Z)\in \mathbb{R}^3$ ($i,j,\dots\!=1,2,3$) are the adiabatic parameters, which are assumed to satisfy $XZ-Y^{2}>0$. The transformation from the variables $(q,p)$ to the action-angle variables $(\varphi,I)$ is well known and turns out to be
\begin{subequations}
\begin{eqnarray}
&&q(\varphi,I;x)=\left(\frac{2ZI}{\omega}\right)^{1/2}\sin\varphi,\label{gho:q}\\
&&p(\varphi,I;x)=\left( \frac{2ZI}{\omega} \right)^{1/2} \left(-\frac{Y}{Z}\sin\varphi+\frac{\omega}{Z}\cos\varphi\right),\label{gho:p}
\end{eqnarray}
\end{subequations}
where $\omega:=(XZ-Y^{2})^{1/2}$ is the parameter-dependent angular frequency. Moreover, the generating function of this transformation, in terms of action-angle variables, is
\begin{equation}\label{gho:S}
S(\varphi,I;x)=-\frac{YI}{\omega}\sin^{2}\varphi+I(\varphi+\sin\varphi\cos\varphi).
\end{equation}
Then, putting Eqs.~(\ref{gho:q}), (\ref{gho:p}), and (\ref{gho:S}) into Eq.~(\ref{classical:G2}), we obtain the functions $G_i(\varphi,I;x)$:
\begin{subequations}
	\begin{eqnarray}
	&&G_{1}(\varphi,I;x)=-\frac{ZI}{2\omega^{2}}\sin\varphi\cos\varphi, \label{gho:G1}\\
	&&G_{2}(\varphi,I;x)=\frac{I\sin\varphi}{\omega^{2}}\left(Y\cos\varphi+\omega\sin\varphi\right),\label{gho:G2}\\
	&&G_{3}(\varphi,I;x)=\frac{I\sin\varphi}{2Z\omega^{2}}\left[(XZ-2Y^{2})\cos\varphi-2Y\omega\sin\varphi\right].\label{gho:G3}\nonumber\\
	\end{eqnarray}
\end{subequations}
Rewriting these functions in terms of the variables $(q,p)$, they satisfy Eqs.~(\ref{classical:Gq}) and (\ref{classical:Gp}).  With this at hand, Eq.~(\ref{gclas:metric}) can now be readily applied to Eqs.~(\ref{gho:G1}), (\ref{gho:G2}), and (\ref{gho:G3}). This yields the components of the corresponding classical metric $g_{ij}(I;x)$, which can be expressed as
\begin{equation}\label{gho:classmetric}
g_{ij}(I;x)=\frac{I^{2}}{32\omega^{4}}\left(\begin{array}{ccc}
Z^{2} & -2YZ & 2Y^{2}-XZ\\
-2YZ & 4XZ & -2XY\\
2Y^{2}-XZ & -2XY & X^{2}
\end{array}\right).
\end{equation}

The idea is now to compare this metric with that coming from the quantum metric tensor~(\ref{QIM}). In the quantum case, the time-dependent Hamiltonian operator $\hat{H}$ of the system is
\begin{equation}\label{gho:quantumH}
\hat{H}=\frac{1}{2}\left[X\hat{q}^{2}+Y(\hat{q}\hat{p}+\hat{p}\hat{q})+Z\hat{p}^{2}\right],
\end{equation}
and leads to the Schr\"odinger equation (with fixed parameters)
\begin{equation}
-\frac{Z\hbar^{2}}{2}\frac{d^{2}\psi_{n}}{dq^{2}}-i\hbar Yq\frac{d\psi_{n}}{dq}+\left(\frac{Xq^{2}}{2}-i\hbar\frac{Y}{2}\right)\psi_{n}=E_{n}\psi_{n},
\end{equation}
which has the normalized solution 
\begin{equation}\label{gho:wavefunctgen}
\psi_{n}(q;x)=\left(\frac{\omega}{Z\hbar}\right)^{1/4}\chi_{n}\left(q\sqrt{\frac{\omega}{Z\hbar}}\right)\exp\left(-\frac{iYq^{2}}{2Z\hbar}\right),
\end{equation}
where $\omega:=(XZ-Y^{2})^{1/2}$ which implies $XZ-Y^{2}>0$,  and $\chi_{n}(\xi)=\left(2^{n}n!\sqrt{\pi}\right)^{-1/2}e^{-\xi^{2}/2}H_{n}(\xi)$ are the Hermite functions, with $H_{n}(\xi)=(-1)^n e^{\xi^2} \frac{d^n}{d\xi^n}e^{-\xi^2}$ being the Hermite polynomials. Furthermore, the energy eigenvalues are given by $E_{n}=(n+1/2)\hbar\omega$ where $n$ are nonnegative integers.

Substituting the wave function (\ref{gho:wavefunctgen}) into $\braket{n}{\partial_in}=\int_{-\infty}^{\infty}dq\psi_{n}^{*}(q,x) \partial_i\psi_{n}(q,x) $ and $\braket{\partial_i n}{\partial_j n}=\int_{-\infty}^{\infty}dq\partial_i\psi_{n}^{*}(q,x) \partial_j\psi_{n}(q,x)$, and bearing in mind the following properties of the Hermite functions
\begin{eqnarray}\label{Hermite}
&&\int_{-\infty}^{\infty}d\xi \ \chi_{m}(\xi)\chi_{n}(\xi)=\delta_{mn},\nonumber\\
&&\frac{d }{d\xi} \chi_{n} = \sqrt{\frac{n}{2}} \ \chi_{n-1} - \sqrt{\frac{n+1}{2}} \ \chi_{n+1}, \nonumber\\
&&\xi \, \chi_{n} = \sqrt{\frac{n}{2}} \ \chi_{n-1} + \sqrt{\frac{n+1}{2}} \ \chi_{n+1},
\end{eqnarray}
the components of the quantum metric~(\ref{QIM}) become
\begin{equation}\label{gho:QIM}
g^{(n)}_{ij}(x)\!=\!\frac{n^{2}\!+\!n\!+\!1}{32\omega^{4}}\!\left(\!\begin{array}{ccc}
Z^{2} & -2YZ & 2Y^{2}-XZ\\
-2YZ & 4XZ & -2XY\\
2Y^{2}-XZ & -2XY & X^{2}
\end{array}\!\right).
\end{equation}

Comparing the metrics (\ref{gho:classmetric}) and (\ref{gho:QIM}), it is clear that they are related  as follows:
\begin{equation}\label{gho:relation}
g^{(n)}_{ij}(x) = \gamma\, g_{ij}(I;x),
\end{equation}
where 
\begin{equation}
\gamma:=\frac{n^{2}+n+1}{I^{2}}.
\end{equation}
Therefore, for the generalized harmonic oscillator, the quantum metric tensor $g^{(n)}_{ij}(x)$ can be determined from the classical metric $g_{ij}(I;x)$, modulo the parameter-independent constant factor~$\gamma$.  This result is nontrivial and supports our claim that the metric (\ref{gclas:metric}) is the classical counterpart of the metric (\ref{QIM}). Note that if we take into account the Bohr-Sommerfeld quantization rule for action variable
\begin{equation}\label{gho:action}
I=\left(n+\frac{1}{2}\right)\hbar,
\end{equation}
then $\gamma $ turns out to be proportional to~$1/\hbar^2$. Additionally, by using Eq.~(\ref{gho:action}), the metrics  (\ref{gho:classmetric}) and (\ref{gho:QIM}) can also be related by 
\begin{equation}\label{gho:relation2}
\frac{\partial}{\partial n} g^{(n)}_{ij}(x) = \frac{1}{\hbar} \frac{\partial}{\partial I} g_{ij}(I;x).
\end{equation}

On another hand, it is worth pointing out that the determinants of the metrics (\ref{gho:classmetric}) and (\ref{gho:QIM}) are zero, which indicates that the corresponding Hamiltonians (\ref{gho:classicalH}) and (\ref{gho:quantumH}) involve more parameters than the effective ones. Actually, these metrics have rank two and hence, to have metrics with nonvanishing determinants, we must leave one of the parameters fixed (but different from zero). In this case, the degeneration in the metric only indicates that one of the parameters is redundant and can be set equal to a constant. To show this explicitly, let us consider for a moment that $\{y^{i'}\}=(X,Y)$ ($i',j',\dots\!=1,2$) are the adiabatic parameters and suppose that $Z=Z_0$ is a nonvanishing constant. In this case, the classical metric (\ref{gclas:metric}) becomes 
\begin{equation}\label{gho:XY}
g_{i'j'}(I;y)=\frac{Z_0 I^{2}}{32\omega^{4}}\left(\begin{array}{cc}
Z_0 & -2Y \\
-2Y & 4X  
\end{array}\right),
\end{equation}
and its determinant, $\det\left[g_{i'j'}(I;y) \right]= \frac{Z_0^2 I^4}{256 \omega^{6} }$, is not zero. Besides, the corresponding quantum metric tensor $g^{(n)}_{i'j'}(y)$, which can be obtained from Eq.~(\ref{gho:XY}) by replacing $ I^{2}$ with $n^{2}+n+1$, also has
a nonvanishing determinant.

Before concluding this example, it may be interesting to obtain the components of the connection and curvature associated with Hannay's angle through Eqs.~(\ref{gclas:connection}) and (\ref{gclas:curvature}), respectively. Using the functions $G_i$ given by Eqs.~(\ref{gho:G1}), (\ref{gho:G2}), and (\ref{gho:G3}), the components of  connection (\ref{gclas:connection}) lead to
\begin{equation}\label{gho:Hconnetion}
A_{1}(I;x)=0,\qquad A_{2}(I;x)=\frac{I}{2\omega},\qquad A_{3}(I;x)=-\frac{YI}{2Z\omega},
\end{equation}
whereas the components of the curvature (\ref{gclas:curvature})  give
\begin{eqnarray}\label{gho:Hcurvature}
&&F_{12}(I;x)=-\frac{ZI}{4\omega^{3}},\qquad
F_{13}(I;x)=\frac{YI}{4\omega^{3}},\nonumber \\ 
&&F_{23}(I;x)=-\frac{XI}{4\omega^{3}}.
\end{eqnarray}
It is also instructive to compare Eqs.~(\ref{gho:Hconnetion}) and (\ref{gho:Hcurvature}) with their quantum analogues, namely Berry's connection and its curvature, respectively. Then, using the wave function~(\ref{gho:wavefunctgen}) and Eq.~(\ref{Hermite}), the components $A^{(n)}_i(x) := - {\rm Im} (\braket{n}{\partial_i n})$ of Berry's connection become
\begin{equation}\label{gho:Bconnetion}
A_{1}^{(n)}(x)=0,\qquad A_{2}^{(n)}(x)=\frac{c_n}{2\omega}, \qquad A_{3}^{(n)}(x)=-\frac{c_n Y}{2Z\omega}, 
\end{equation}
and the components $F^{(n)}_{ij}(x):=\partial_i A^{(n)}_j-\partial_j A^{(n)}_i$ of its curvature yield
\begin{eqnarray}\label{gho:Bcurvature}
&&F_{12}^{(n)}(x)=-\frac{c_n Z}{4\omega^{3}},\qquad 
F_{13}^{(n)}(x)=\frac{c_nY}{4\omega^{3}}, \nonumber\\
&&F_{23}^{(n)}(x)=-\frac{c_n X}{4\omega^{3}}, 
\end{eqnarray}
where $c_n:=\left(n+1/2\right)$. Comparing Eqs.~(\ref{gho:Hconnetion}) and (\ref{gho:Bconnetion}) as well as Eqs.~(\ref{gho:Hcurvature}) and (\ref{gho:Bcurvature}), it is straightforward to see the following relations:
\begin{subequations}
	\begin{eqnarray}
	A_{i}^{(n)}(x)=\beta \, A_{i}(I;x), \label{gho:Arelation}\\
	F_{ij}^{(n)}(x)=\beta \, F_{ij}(I;x), \label{gho:Frelation}
	\end{eqnarray}
\end{subequations}
where
\begin{equation}\label{gho:beta}
\beta:=\frac{n+\frac{1}{2}}{I}.
\end{equation}
This entails that $A_{i}^{(n)}(x)$ and $F_{ij}^{(n)}(x)$ can be obtained, respectively, from $A_{i}(I;x)$ and $F_{ij}(I;x)$, modulo the parameter-independent  constant factor $\beta$. Moreover, after using Eq.~(\ref{gho:action}), this factor reduces to $\beta=1/\hbar$. 

Some comments are in order. First, it is noteworthy to emphasize that  connection defined by dropping $\left<( \partial_i S )_{\varphi,I}\right>$ from Eq.~(\ref{gclas:connection}), namely $A_i(I;x)=\left<p_a ( \partial_i q^a)_{\varphi,I}\right>$, does not lead to Eqs.~(\ref{gho:Hconnetion}) and  therefore does not satisfy the relation (\ref{gho:Arelation}). Of course, the curvature of such a connection, which is also given by Eq.~(\ref{gclas:Curv}), implies Eq.~(\ref{gho:Hcurvature}). Second, notice that Eq.~(\ref{gho:Frelation}) is in complete agreement with the semiclassical relation between Berry's curvature and the curvature associated with Hannay's angle reported in Ref.~\cite{Berry1985}. Finally, it is worth mentioning that the multiplicative constants involved in the relation (\ref{gho:relation}) and the relations (\ref{gho:Arelation}) and (\ref{gho:Frelation}) are different: while $\gamma$ is proportional to $1/I^2$, $\beta$ is proportional to $1/I$.

\subsection{Generalized harmonic oscillator with a linear term}

For our second example we shall consider the generalized harmonic oscillator with a linear term in the position. Thus the Hamiltonian under consideration is
\begin{equation}\label{ghoWq:classicalH}
H=\frac{1}{2}\left(X q^2+2Yqp+Zp^2\right)+Wq,
\end{equation}
where $x=\{x^{i}\}=(W,X,Y,Z)$ with $i,j,\dots\!=0,1,2,3$  are the adiabatic parameters. Assuming $XZ-Y^{2}>0$, we find that the variables $(q,p)$ in terms of  action-angle variables $(\varphi,I)$ read
\begin{subequations}
	\begin{eqnarray}
	&&q(\varphi,I;x)=\left(\frac{2ZI}{\omega}\right)^{1/2}\sin\varphi-\frac{WZ}{\omega^{2}},\label{ghoWq:q}\\
	&&p(\varphi,I;x)=\left( \frac{2ZI}{\omega} \right)^{1/2} \left(-\frac{Y}{Z}\sin\varphi+\frac{\omega}{Z}\cos\varphi\right)+\frac{WY}{\omega^{2}},\nonumber\\\label{ghoWq:p}
	\end{eqnarray}
\end{subequations}
where $\omega:=(XZ-Y^{2})^{1/2}$ is the angular frequency of the system, which is independent of $W$. Furthermore, we get that the generating function $S(\varphi,I;x)$ of the transformation $(q,p) \rightarrow (\varphi,I)$ is
\begin{eqnarray}\label{ghoWq:S}
S(\varphi,I;x)&=&-\frac{Y}{2Z}\left[\left(\frac{2ZI}{\omega}\right)^{1/2} \sin\varphi-\frac{WZ}{\omega^{2}}\right]^{2} \nonumber\\
&&+I(\varphi+\sin\varphi\cos\varphi).
\end{eqnarray}
With these ingredients at hand,  it is straightforward to obtain  $G_i(\varphi,I;x)$ from Eq.~(\ref{classical:G2}). The resulting functions, in compact form, are 
\begin{equation}\label{ghoWq:Gi}
G_i(\varphi,I;x)=f_i(x) qp +g_i(x) q^2 + h_i(x) \left(p + \frac{Y}{Z} q \right) ,
\end{equation}
where  $p=p(\varphi,I;x)$ and $q=q(\varphi,I;x)$ are given by Eqs.~(\ref{ghoWq:q}) and (\ref{ghoWq:p}), respectively, while
\begin{subequations}
	\begin{eqnarray}
	&&f_i(x):= \frac{\omega}{2Z} \, \partial_i\left( \frac{Z}{\omega}\right), \label{ghoWq:f}\\
	&&g_i(x):= \frac{Y}{Z} f_i(x)  + \frac{1}{2} \, \partial_i \left(\frac{Y}{Z}\right), \label{ghoWq:g}\\
	&&h_i(x):= \frac{W}{2\omega} \, \partial_i  \left( \frac{Z}{\omega}\right) - \partial_i \left( \frac{WZ}{\omega^2}\right).\label{ghoWq:h}
	\end{eqnarray}
\end{subequations}
It can be verified that $G_i$ given by Eq.~(\ref{ghoWq:Gi}) satisfy Eqs.~(\ref{classical:Gq}) and (\ref{classical:Gp}). In addition, if the parameter $W$ is fixed to 0, it is not difficult to realize that these functions reduce to those of Eqs.~(\ref{gho:G1}), (\ref{gho:G2}), and~(\ref{gho:G3}). 

By inserting Eq.~(\ref{ghoWq:Gi}) into Eq.~(\ref{gclas:metric}), the corresponding  components of the classical metric $g_{ij}(I;x)$ are
\begin{widetext}
\begin{multline}\label{ghoWq:classmetric}
g_{ij}(I;x)=\frac{I^{2}}{32\omega^{4}}\begin{pmatrix}0 & 0 & 0 & 0\\
0 & Z^{2} & -2YZ & 2Y^{2}-XZ\\
0 & -2YZ & 4XZ & -2XY\\
0 & 2Y^{2}-XZ & -2XY & X^{2}
\end{pmatrix}\\
+\frac{I}{\omega^{7}}\begin{pmatrix}Z\omega^{4} & -WZ^{2}\omega^{2} & 2WYZ\omega^{2} & -WY^{2}\omega^{2}\\
-WZ^{2}\omega^{2} & W^{2}Z^{3} & -2W^{2}YZ^{2} & W^{2}Y^{2}Z\\
2WYZ\omega^{2} & -2W^{2}YZ^{2} & W^{2}Z(3Y^{2}+XZ) & -W^{2}Y(Y^{2}+XZ)\\
-WY^{2}\omega^{2} & W^{2}Y^{2}Z & -W^{2}Y(Y^{2}+XZ) & W^{2}XY^{2}
\end{pmatrix}.
\end{multline}
\end{widetext}
Notice  that this metric has an extra term, as compared to the metric~(\ref{gho:classmetric}), which is proportional to $I/\omega^{7}$ and is a consequence of the linear modification introduced in the Hamiltonian~(\ref{ghoWq:classicalH}). Certainly, by fixing $W=0$ in the above expression and eliminating the corresponding row and column, we can recover the metric (\ref{gho:classmetric}).

To contrast Eq.~(\ref{ghoWq:classmetric}) with the quantum metric tensor, we consider the following  Hamiltonian operator:
\begin{equation}\label{ghoWq:quantumH}
\hat{H}=\frac{1}{2}\left[X\hat{q}^{2}+Y(\hat{q}\hat{p}+\hat{p}\hat{q})+Z\hat{p}^{2}\right]+W\hat{q}.
\end{equation}
In this case the Schr\"odinger equation reads
\begin{equation}
-\frac{Z\hbar^{2}}{2}\frac{d^{2}\psi_{n}}{dq^{2}}-i\hbar Yq\frac{d\psi_{n}}{dq}+\left(\frac{Xq^{2}}{2}\!+\!Wq\!-\! i\hbar\frac{Y}{2}\right)\!\psi_{n}\!=\! E_{n}\psi_{n},
\end{equation}
and the eigenfunctions $\psi_{n}(q;x)$ are of the form
\begin{eqnarray}\label{ghoWq:wavefunctgen}
&&\psi_{n}(q;x)\equiv\nonumber\\ &&\left(\frac{\omega}{Z\hbar}\right)^{1/4}\chi_{n}\! \left[\left(q+\frac{WZ}{\omega^{2}}\right)\sqrt{\frac{\omega}{Z\hbar}}\right]\exp\!\left(-\frac{iYq^{2}}{2Z\hbar}\right),
\end{eqnarray}
where once again $\omega=(XZ-Y^{2})^{1/2}$ which entails $XZ-Y^{2}>0$. By substituting Eq.~(\ref{ghoWq:wavefunctgen}) into Eq.~(\ref{QIM}) and using Eq.~(\ref{Hermite}), we get that the components of the quantum metric $g^{(n)}_{ij}(x)$ are given by
\begin{widetext}
\begin{multline}\label{ghoWq:QIM}
g^{(n)}_{ij}(x)=\frac{n^{2}+n+1}{32\omega^{4}}\begin{pmatrix}0 & 0 & 0 & 0\\
0 & Z^{2} & -2YZ & 2Y^{2}-XZ\\
0 & -2YZ & 4XZ & -2XY\\
0 & 2Y^{2}-XZ & -2XY & X^{2}
\end{pmatrix}\\
+\frac{n+\frac{1}{2}}{\hbar\omega^{7}}\begin{pmatrix}Z\omega^{4} & -WZ^{2}\omega^{2} & 2WYZ\omega^{2} & -WY^{2}\omega^{2}\\
-WZ^{2}\omega^{2} & W^{2}Z^{3} & -2W^{2}YZ^{2} & W^{2}Y^{2}Z\\
2WYZ\omega^{2} & -2W^{2}YZ^{2} & W^{2}Z(3Y^{2}+XZ) & -W^{2}Y(Y^{2}+XZ)\\
-WY^{2}\omega^{2} & W^{2}Y^{2}Z & -W^{2}Y(Y^{2}+XZ) & W^{2}XY^{2}
\end{pmatrix}.
\end{multline}
\end{widetext}

We can see that the classical metric~(\ref{ghoWq:classmetric}) and the quantum metric (\ref{ghoWq:QIM}) have exactly the same functional dependence on the adiabatic parameters. Hence we corroborate once again that the metric~(\ref{gclas:metric}) is the classical analog of the quantum metric tensor~(\ref{QIM}). Remarkably, by using the Bohr-Sommerfeld quantization rule~(\ref{gho:action}), it follows that the relation~(\ref{gho:relation2}) also holds for the metrics (\ref{ghoWq:classmetric}) and~(\ref{ghoWq:QIM}).

Note that in this example, as well as in the previous one, the metrics $g_{ij}(I;x)$ and $g^{(n)}_{ij}(x)$  have vanishing determinant. However, here the rank of the metrics (\ref{ghoWq:classmetric}) and  (\ref{ghoWq:QIM}) is three, which shows the existence of a redundant parameter. In particular, if we take $\{y^{i'}\}=(W,X,Y)$ ($i',j',\dots\!=0,1,2$) as the adiabatic parameters and $Z=Z_0$ as a nonvanishing constant, then the classical metric reads
\begin{multline}\label{ghoWq:WXY}
	g_{i'j'}(I;y)=\frac{Z_0 I^{2}}{32\omega^{4}}\begin{pmatrix}0 & 0 & 0 \\
	0 & Z_0 & -2Y \\
	0 & -2Y & 4X 
	\end{pmatrix}\\
	+\frac{Z_0 I}{\omega^{7}}\begin{pmatrix}\omega^{4} & -W^{2}Z_0\omega^{2} & 2WY\omega^{2} \\
	-WZ_0\omega^{2} & W^{2}Z_0^{2} & -2W^{2}YZ_0 \\
	2WY\omega^{2} & -2W^{2}YZ_0 & W^{2}(3Y^{2}+XZ_0) 
	\end{pmatrix},
\end{multline}
and its determinant $\det[g_{i'j'}(I;y)]=\frac{Z_0^3 I^4}{256 \omega^{12}} (I \omega^{3}+8W^2 Z_0)$ is different from zero.

To conclude this example, let us obtain the corresponding classical and quantum connections and curvatures. Classically, by applying Eqs.~(\ref{gclas:connection}) and (\ref{gclas:curvature}) to the functions $G_i$ given by Eq.~(\ref{ghoWq:Gi}),  we obtain the components of the connection,
\begin{eqnarray}\label{ghoWq:Hconnetion}
&&A_{0}(I;x)=A_{1}(I;x)=0,\quad A_{2}(I;x)=\frac{I}{2\omega}+\frac{W^{2}Z}{2\omega^{4}}, \nonumber\\ &&A_{3}(I;x)=-\frac{YI}{2Z\omega}-\frac{W^{2}Y}{2\omega^{4}},
\end{eqnarray}
and the components of the curvature, which are displayed in matrix form,
\begin{eqnarray}\label{ghoWq:Hcurvature}
&& F_{ij}(I;x)=\frac{I}{4\omega^{3}}\begin{pmatrix}0 & 0 & 0 & 0\\
0 & 0 & -Z & Y\\
0 & Z & 0 & -X\\
0 & -Y & X & 0
\end{pmatrix} \nonumber\\
&&+\frac{1}{\omega^{6}}\begin{pmatrix}0 & 0 & WZ\omega^{2} & -WY\omega^{2}\\
0 & 0 & -W^{2}Z^{2} & W^{2}YZ\\
-WZ\omega^{2} & W^{2}Z^{2} & 0 & -W^{2}Y^{2}\\
WY\omega^{2} & -W^{2}YZ & W^{2}Y^{2} & 0
\end{pmatrix} \!,
\end{eqnarray}
respectively. On the quantum side,  Berry's connection and curvature obtained from the eigenfunctions~(\ref{ghoWq:wavefunctgen}) are
\begin{eqnarray}\label{ghoWq:Bconnetion}
&&A_{0}^{(n)}(x)=A_{1}^{(n)}(x)=0,\qquad A_{2}^{(n)}(x)=\frac{n+\frac{1}{2}}{2\omega}+\frac{W^{2}Z}{2\hbar\omega^{4}},\nonumber\\ &&A_{3}^{(n)}(x)=-\frac{\left(n+\frac{1}{2}\right)Y}{2Z\omega}-\frac{W^{2}Y}{2\hbar\omega^{4}},
\end{eqnarray}
and
\begin{eqnarray}\label{ghoWq:Bcurvature}
&&F^{(n)}_{ij}(x)=\frac{n+\frac{1}{2}}{4\omega^{3}}\begin{pmatrix}0 & 0 & 0 & 0\\
0 & 0 & -Z & Y\\
0 & Z & 0 & -X\\
0 & -Y & X & 0
\end{pmatrix} \nonumber \\
&& +\frac{1}{\hbar\omega^{6}} \! \begin{pmatrix}0 & 0 & WZ\omega^{2} & -WY\omega^{2}\\
0 & 0 & -W^{2}Z^{2} & W^{2}YZ\\
-WZ\omega^{2} & W^{2}Z^{2} & 0 & -W^{2}Y^{2}\\
WY\omega^{2} & -W^{2}YZ & W^{2}Y^{2} & 0
\end{pmatrix}, \
\end{eqnarray}
respectively. Here we used once again Eq.~(\ref{Hermite}). By comparing Eqs.~(\ref{ghoWq:Hconnetion}) and~(\ref{ghoWq:Bconnetion}) as well as Eqs.~(\ref{ghoWq:Hcurvature}) and~(\ref{ghoWq:Bcurvature}), it turns out that the relations (\ref{gho:Arelation}) and (\ref{gho:Frelation}) hold provided that the Bohr-Sommerfeld quantization rule~(\ref{gho:action}) is taken into account, i.e., when $\beta=1/\hbar$ in Eq.~(\ref{gho:beta}).

\subsection{Quartic anharmonic oscillator}

In this example, we focus on the classical quartic anharmonic oscillator which is defined by the Hamiltonian
\begin{eqnarray}\label{quartic:Hamiltonial}
H=\frac{p^{2}}{2m}+\frac{k}{2}q^{2}+ \frac{\lambda}{4!}q^{4},
\end{eqnarray}
where $x=\{x^{i}\}=(m,k,\lambda)$ with $i=1,2,3$ are the adiabatic parameters. In this case, in contrast to the previous examples, we need to resort to the canonical perturbation theory in order to find the functions $G_i$. With this in mind, the starting point is to decompose the Hamiltonian~(\ref{quartic:Hamiltonial}) in the form $H=H_{0}+\lambda H_{1}$ where 
\begin{equation}\label{quartic:H0H1}
H_{0}=\frac{p^{2}}{2m}+\frac{k}{2}q^{2}, \qquad H_{1}=\frac{q^{4}}{4\text{!}},
\end{equation}
and we assume $0\leq\lambda\ll1$. Here, $H_{0}$ is the Hamiltonian of the unperturbed problem, for which the action-angle variables $(\varphi_{0},I_{0})$ are well known and allow us to express the variables $(q,p)$ as
\begin{eqnarray}
q(\varphi_{0},I_{0};x)&=&\left(\frac{2I_{0}}{m\omega_{0}}\right)^{1/2} \sin\varphi_{0}, \label{quartic:q}\\
p(\varphi_{0},I_{0};x)&=&\left(2m\omega_{0}I_{0} \right)^{1/2}\cos\varphi_{0},\label{quartic:p}
\end{eqnarray}
where $\omega_{0}=\left(k/m\right)^{1/2}$ is the unperturbed frequency. Furthermore, $H_{1}$ is the perturbative potential. 

Next we assume that the type 2 generating function  $W(\varphi_{0},I;x)$ of the canonical transformation from $(\varphi_{0},I_{0})$ to the action-angle variables $(\varphi,I)$ of the total problem $H(I;x)$ can be expanded in a power series of $\lambda$:
\begin{eqnarray}
W(\varphi_{0},I;x)&=&\varphi_{0}I+\lambda W_{1}(\varphi_{0},I;x)+\lambda^{2}W_{2}(\varphi_{0},I;x)\nonumber\\
&&+\lambda^{3}W_{3}(\varphi_{0},I;x)+{\cal O}(\lambda^{4}),
\end{eqnarray}
where $W_{1},W_{2},\dots,$ are functions to be determined. Thus, the equations of the canonical transformation, $\varphi(\varphi_{0},I;x)=\partial W(\varphi_{0},I;x)/\partial I$ and $I_{0}(\varphi_{0},I;x)=\partial W(\varphi_{0},I;x)/\partial \varphi_{0}$, take the form
\begin{eqnarray}
\varphi(\varphi_{0},I;x)&=&\varphi_{0}+\lambda\frac{\partial W_{1}(\varphi_{0},I;x)}{\partial I}+\lambda^{2}\frac{\partial W_{2}(\varphi_{0},I;x)}{\partial I}\nonumber\\
&&+\lambda^{3}\frac{\partial W_{3}(\varphi_{0},I;x)}{\partial I}+{\cal O}(\lambda^{4})\label{quartic:varphi0},
\end{eqnarray}
and
\begin{eqnarray}
I_{0}(\varphi_{0},I;x)&=&I+\lambda\frac{\partial W_{1}(\varphi_{0},I;x)}{\partial\varphi_{0}}+\lambda^{2}\frac{\partial W_{2}(\varphi_{0},I;x)}{\partial\varphi_{0}}\nonumber\\
&&+\lambda^{3}\frac{\partial W_{3}(\varphi_{0},I;x)}{\partial\varphi_{0}}+{\cal O}(\lambda^{4}), \label{quartic:I0}
\end{eqnarray}
respectively.

Following the canonical perturbation theory and working up to the third order in $\lambda$, the functions  $W_{1}$, $W_{2}$, and $W_{3}$  can be obtained by solving the differential equations~\cite{goldstein2000,dittrich2017}
\begin{equation}\label{quartic:diffW}
\omega_{0}\frac{\partial W_{\mu}(\varphi_{0},I;x)}{\partial\varphi_{0}}=\left<\Phi_{\mu}(\varphi_{0},I;x)\right>_0-\Phi_{\mu}(\varphi_{0},I;x), 
\end{equation}
where $\left< \cdot \right>_{0}$  denotes the average with respect to $\varphi_{0}$ and, in our case, $\Phi_{1}=H_{1}$, $\Phi_{2}=\frac{\partial W_{1}}{\partial\varphi_{0}} \frac{\partial H_{1}}{\partial I}$ and $\Phi_{3}=\frac{1}{2}\left(\frac{\partial W_{1}}{\partial\varphi_{0}}\right)^2\frac{\partial^2 H_{1}}{\partial I^2}+\frac{\partial W_{2}}{\partial\varphi_{0}}\frac{\partial H_{1}}{\partial I}$. Explicitly these functions are given by 
\begin{subequations}
\begin{eqnarray}
&&\Phi_{1}(\varphi_{0},I;x)=\frac{I^{2}\sin^{4}\varphi_{0}}{6m^{2}\omega_{0}^{2}},\\
&&\Phi_{2}(\varphi_{0},I;x)=-\frac{I^{3}\sin^{4}\varphi_{0}}{144m^{4}\omega_{0}^{5}}\left(8\sin^{4}\varphi_{0}-3\right),\\
&&\Phi_{3}(\varphi_{0},I;x)=\frac{I^{4}\sin^{4}\varphi_{0}}{13824m^{6}\omega_{0}^{8}}\left(320\sin^{8}\varphi_{0}-144\sin^{4}\varphi_{0}-25\right), \nonumber\\
\end{eqnarray}
\end{subequations}
and together with Eq.~(\ref{quartic:diffW}) they imply
\begin{subequations}
\begin{eqnarray}
&&W_{1}(\varphi_{0},I;x)=\frac{I^{2}}{192m^{2}\omega_{0}^{3}}(8\sin2\varphi_{0}-\sin4\varphi_{0}),\\
&&W_{2}(\varphi_{0},I;x)=\frac{I^{3}}{55296m^{4}\omega_{0}^{6}} \left(-384\sin2\varphi_{0}+132\sin4\varphi_{0}\right.\nonumber\\
&&\left.-32\sin6\varphi_{0}+3\sin8\varphi_{0}\right),\\
&&W_{3}(\varphi_{0},I;x)=\frac{I^{4}}{5308416m^{6}\omega_{0}^{9}}(9264\sin2\varphi_{0}-4101\sin4\varphi_{0}\nonumber\\
&&+1624\sin6\varphi_{0}\!-\!441\sin8\varphi_{0}\!+\!72\sin10\varphi_{0}\!-\!5\sin12\varphi_{0}). \ \ \ \ \ \
\end{eqnarray}
\end{subequations}

Having found $W(\varphi_{0},I;x)$, it is straightforward to obtain the generating function $S(\varphi,I;x)$ of the canonical transformation $(q,p) \rightarrow (\varphi,I)$. Indeed, since the transformation $(q,p) \rightarrow (\varphi,I)$ can be regarded as the composition of the successive canonical transformations $(q,p)\rightarrow(\varphi_{0},I_{0})$  and $(\varphi_{0},I_{0})\rightarrow(\varphi,I)$,  generated, respectively, by $S_{0}(q,I_{0};x)$ and $W(\varphi_{0},I;x)$, we have that  $S(\varphi,I;x)$ is given by
\begin{equation} \label{quartic:S}
S(q,I;x)=S_{0}(q,I_{0};x)+W(\varphi_{0},I;x)-\varphi_{0}I_{0},
\end{equation}
where the function $S_{0}$ in terms of the variables $(\varphi_{0},I_{0})$ reads
\begin{equation}\label{quartic:S0}
S_{0}(\varphi_{0},I_{0};x)=I_{0}(\varphi_{0}+\sin\varphi_{0}\cos\varphi_{0}).
\end{equation}

Note that substituting Eq.~(\ref{quartic:I0}) into Eqs.~(\ref{quartic:q}), (\ref{quartic:p}), and (\ref{quartic:S}), we can write the variables $q$ and $p$ and the function $S$ in terms of $\varphi_{0}$ and~$I$. Now, to compute the metric~(\ref{gclas:metric}) we need to first obtain the functions  $G_{i}(\varphi,I;x)=p(\partial_{i}q)_{\varphi,I}-(\partial_{i}S)_{\varphi,I}$, where the derivatives are taken at fixed action-angle variables $(\varphi,I)$. These derivatives can be achieved by employing the following useful formula:
\begin{equation}\label{quartic:formula}
\left(\partial_i \mathcal{F}\right)_{\varphi,I}=\left(\partial_i \mathcal{F}\right)_{\varphi_{0},I}-\frac{(\partial \mathcal{F}/\partial\varphi_{0})_{I,x}}{(\partial\varphi/\partial\varphi_{0})_{I,x}}\left(\partial_i \varphi\right)_{\varphi_{0},I},
\end{equation}
where $\mathcal{F}(\varphi_{0},I;x)=\mathcal{F}(\varphi(\varphi_{0},I;x),I;x)$ is  $q(\varphi_{0},I;x)$ or  $S(\varphi_{0},I;x)$.  Notice that we can use Eq.~(\ref{quartic:varphi0}) to compute $(\partial\varphi/\partial\varphi_{0})_{I,x}$ and $\left(\partial_i \varphi\right)_{\varphi_{0},I}$. After carrying out these calculations and  retaining terms correct to second order in $\lambda$ (since derivatives with respect to this parameter are involved), we arrive at the functions $G_i$ in terms of $\varphi_{0}$ and~$I$, namely
\begin{equation}\label{quartic:G1}
G_i(\varphi_{0},I;x)=\alpha_{i0}+\alpha_{i1} \lambda +\alpha_{i2} \lambda^2,
\end{equation}
where in the case $i=1$:
\begin{subequations}
	\begin{eqnarray}
	\alpha_{10}&=&-\frac{I \sin 2 \varphi _0}{4 m},\\
	\alpha_{11}&=&-\frac{I^2  \sin^3\varphi _0}{48 \sqrt{k^3 m^3}}\left(\cos 3 \varphi _0-2 \cos \varphi _0\right),\\
	\alpha_{12}&=&-\frac{I^3 }{55296 k^3 m^2} \left(318 \sin 2 \varphi _0-204 \sin 4 \varphi _0\right.\nonumber\\
	&&\left.+95 \sin 6 \varphi _0-27 \sin 8 \varphi _0+3 \sin 10 \varphi _0\right),
	\end{eqnarray}
\end{subequations}
in the case $i=2$:
\begin{subequations}
	\begin{eqnarray}
	\alpha_{20}&=&-\frac{I \sin 2 \varphi _0}{4 k},\\
	\alpha_{21}&=&\frac{I^2}{384 \sqrt{k^5 m}}\left(23 \sin 2 \varphi _0-7 \sin 4 \varphi _0+\sin 6 \varphi _0\right), \ \ \ \ \ \\
	\alpha_{22}&=&-\frac{I^3 }{18432 k^4 m}\left(362 \sin 2 \varphi _0-156 \sin 4 \varphi _0\right.\nonumber\\
	&&\left.+53 \sin 6 \varphi _0-11 \sin 8 \varphi _0+\sin 10 \varphi _0\right),
	\end{eqnarray}
\end{subequations}
and in the case $i=3$:
\begin{subequations}
	\begin{eqnarray}
	&&\alpha_{30}=\frac{I^2 }{192 \sqrt{k^3 m}}\left(\sin 4 \varphi _0-8 \sin 2 \varphi _0\right)\\
	&&\alpha_{31}=\frac{I^3}{27648 k^3 m}\left(384 \sin 2 \varphi _0-132 \sin 4 \varphi _0\right.\nonumber\\
	&&\left.\hspace{9mm}+32 \sin 6 \varphi _0-3 \sin 8 \varphi _0\right),\\
	&&\alpha_{32}=\frac{I^4 }{1769472 \sqrt{k^{9} m^{3}}} \left(-9264 \sin 2 \varphi _0+4101 \sin 4 \varphi _0\right.\nonumber\\
	&&\left.-1624 \sin 6 \varphi _0+441 \sin 8 \varphi _0-72 \sin 10 \varphi _0+5 \sin 12 \varphi _0\right).\nonumber\\
	\end{eqnarray}
\end{subequations}

Finally, substituting $G_i(\varphi_{0},I;x)$ into Eq.~(\ref{gclas:metric}) and writing the average over the angle variable~$\varphi$ as
\begin{equation}
\left< f(\varphi)  \right>\!=\!\frac{1}{2 \pi}\!\int_0^{2 \pi} \! d\varphi f(\varphi)\!=\!\frac{1}{2 \pi}\! \int_0^{2 \pi} \! d\varphi_0 \! \left(\frac{\partial\varphi}{\partial\varphi_{0}}\right)_{I,x} f(\varphi_0),\nonumber
\end{equation}
we obtain the components of the classical metric $g_{ij}(I;x)$ correct to second order in $\lambda$:
\begin{eqnarray}\label{quartic:classmetric}
&&g_{11}(I;x)=\frac{I^{2}}{32m^{2}}-\frac{\lambda I^{3}}{256\sqrt{m^{5}k^{3}}}+\frac{47\lambda^{2} I^{4}}{32768m^{3}k^{3}},\nonumber\\
&&g_{12}(I;x)=\frac{I^{2}}{32mk}-\frac{7\lambda I^{3}}{768\sqrt{m^{3}k^{5}}}+\frac{347\lambda^{2} I^{4}}{98304m^{2}k^{4}},\nonumber\\
&&g_{13}(I;x)=\frac{I^{3}}{192\sqrt{m^{3}k^{3}}}-\frac{103\lambda I^{4}}{49152m^{2}k^{3}}+\frac{15\lambda^{2} I^{5}}{16384\sqrt{m^{5}k^{9}}},\nonumber\\
&&g_{22}(I;x)=\frac{I^{2}}{32k^{2}}-\frac{11\lambda I^{3}}{768\sqrt{mk^{7}}}+\frac{1919\lambda^{2}I^{4}}{294912mk^{5}},\nonumber\\
&&g_{23}(I;x)=\frac{I^{3}}{192\sqrt{mk^{5}}}-\frac{439\lambda I^{4}}{147456mk^{4}}+\frac{7\lambda^{2} I^{5}}{4608\sqrt{m^{3}k^{11}}},\nonumber\\
&&g_{33}(I;x)\!=\!\frac{65I^{4}}{73728mk^{3}}\!-\!\frac{89\lambda I^{5}}{147456\sqrt{m^{3}k^{9}}}\!+\!\frac{130621 \lambda^{2} I^{6}}{382205952m^{2}k^{6}}. \nonumber\\
\end{eqnarray}

Now we are interested in contrasting Eq.~(\ref{quartic:classmetric}) with its quantum counterpart. For the ground state of the quantum quartic anharmonic oscillator, by using Eq.~(\ref{QIM})  and following a perturbative treatment, we find the corresponding components of the quantum metric tensor $g^{(0)}_{ij}$ with terms up to second order in $\lambda$ (see the Appendix for the details):
\begin{eqnarray}\label{quartic:QIM}
&&g^{(0)}_{11}(x)=\frac{1}{32m^{2}}-\frac{3\hbar\lambda}{512\sqrt{m^{5}k^{3}}}+\frac{59\hbar^{2}\lambda^{2}}{16384m^{3}k^{3}},\nonumber\\
&&g^{(0)}_{12}(x)=\frac{1}{32mk}-\frac{7\hbar\lambda}{512\sqrt{m^{3}k^{5}}}+\frac{143\hbar^{2}\lambda^{2}}{16384m^{2}k^{4}},\nonumber\\
&&g^{(0)}_{13}(x)=\frac{\hbar}{128\sqrt{m^{3}k^{3}}}-\frac{21\hbar^{2}\lambda}{4096m^{2}k^{3}}+\frac{2353\hbar^{3}\lambda^{2}}{589824\sqrt{m^{5}k^{9}}},\nonumber\\
&&g^{(0)}_{22}(x)=\frac{1}{32k^{2}}-\frac{11\hbar\lambda}{512\sqrt{mk^{7}}}+\frac{785\hbar^{2}\lambda^{2}}{49152mk^{5}},\nonumber\\
&&g^{(0)}_{23}(x)=\frac{\hbar}{128\sqrt{mk^{5}}}-\frac{89\hbar^{2}\lambda}{12288mk^{4}}+\frac{3841\hbar^{3}\lambda^{2}}{589824\sqrt{m^{3}k^{11}}},\nonumber\\
&&g^{(0)}_{33}(x)=\frac{13\hbar^{2}}{6144mk^{3}}\!-\!\frac{31\hbar^{3}\lambda}{12288\sqrt{m^{3}k^{9}}}\!+\!\frac{57227\hbar^{4}\lambda^{2}}{21233664m^{2}k^{6}}. \ \ \ \
\end{eqnarray}

Note that by multiplying the components $g^{(0)}_{ij}$ in Eq.~(\ref{quartic:QIM}) by $\hbar^2$ and comparing the result with the corresponding components $g_{ij}(I;x)$ in Eq.~(\ref{quartic:classmetric}), we have that terms with the same powers of $\hbar$ and $I$ have exactly the same functional dependence on the parameters $m$, $k$, and $\lambda$. Then, to match Eqs.~(\ref{quartic:classmetric}) and~(\ref{quartic:QIM}), it is reasonable to consider $g^{(0)}_{ij}=\frac{1}{\hbar^2} g_{ij}(I;x)$. By doing this, we find the following identifications for the ground state: $I^{2}=\hbar^{2}$, $I^{3}=\frac{3}{2}\hbar^{3}$, and $I^{6}=\frac{1030086}{130621}\hbar^{6}$. In the case of $I^{4}$ and $I^{5}$, the identifications are not unique, but they differ from each other slightly. Actually, for $I^{4}$ we find  $I^4\approx2.4\hbar^{4},2.43\hbar^{4},2.44\hbar^{4},2.45\hbar^{4},2.47\hbar^{4},2.51\hbar^{4}$, whereas for $I^{5}$ we find $I^5\approx4.18\hbar^{5},4.29\hbar^{5},4.35\hbar^{5}$. Therefore,  $g^{(0)}_{ij}$ can only be obtained in an approximate way from $g_{ij}(I;x)$ through the relation $g^{(0)}_{ij}\approx\frac{1}{\hbar^2} g_{ij}(I;x)$ with the appropriate identifications. This result is somehow expected because we have been dealing  with perturbation theories to arrive at Eqs.~(\ref{quartic:classmetric}) and~(\ref{quartic:QIM}). Finally, it is worth mentioning that the determinants of the metrics defined by the components~(\ref{quartic:classmetric}) and~(\ref{quartic:QIM}), obtained by keeping terms up to second order in $\lambda$, are zero.  Nonetheless, if one of the parameters is left fixed, then the corresponding (classical or quantum) metric could have a nonvanishing determinant.

\section{Alternative expressions for the quantum metric tensor and Berry's connection}\label{sec:alternative}

We now extend the use of the classical functions $G_i$ given by Eq.~(\ref{classical:G}) to the quantum case. To this end we start by promoting  the functions $G_i(q, p;x)$, expressed in terms of the variables $(q,p)$, to quantum operators $\hat{G}_i(\hat{q},\hat{p};x)$ which we assume to be Hermitian. By analogy with the classical case where $G_i \delta x^i$ generates a displacement in the parameter space, it is reasonable to consider  $\delta x^i\hat{G}_i(\hat{q}, \hat{p};x)$ as the generator the infinitesimal displacement of states $\ket{n(x)}\rightarrow\ket{n(x')}$, namely  $\ket{n(x')}=\exp(- \frac{\rmi}{\hbar}  \delta x^i\hat{G}_i)\ket{n(x)}$. Then, we can replace the operators $\hat{P}_i$ in Eq.~(\ref{QIMOperator2}) by the operators $\hat{G}_i$, obtaining
\begin{equation}\label{QG:def}
\rmi \hbar \ket{\partial_i n(x)}=\hat{G}_i \ket{n(x)}.
\end{equation}
This allows us to write down the quantum metric tensor~(\ref{QIM}) [or Eq.~(\ref{QIM2})] in terms of  $\hat{G}_i(\hat{q},\hat{p};x)$ as
\begin{equation}\label{QG:QIM}
g^{(n)}_{ij}(x) = \frac{1}{\hbar^2} {\rm Re} \left( \langle \hat{G}_i \hat{G}_j \rangle_n - \langle \hat{G}_i \rangle_n \langle \hat{G}_j\rangle_n \right).
\end{equation}
Note that on account of the Hermiticity of $\hat{G}_i$, the r.h.s of this expression is symmetric.

Similarly, we can recast Berry's connection in terms of the operators $\hat{G}_i(\hat{q},\hat{p};x)$. Indeed, using Eq.~(\ref{QG:def}) and recalling that  the expectation values $\langle \hat{G}_i \rangle_n$ are real (by virtue of the Hermiticity of $\hat{G}_i$), we can rewrite Berry's connection, $A^{(n)}_i(x):= - {\rm Im} (\braket{n}{\partial_i n})$, in the following form:
\begin{equation}\label{QG:A}
A^{(n)}_i(x) = \frac{1}{\hbar} \langle \hat{G}_i \rangle_n.
\end{equation}
Finally, we note that that taking into account Eqs.~(\ref{QG:def}) and~(\ref{QG:A}), the action of the operator $\Delta \hat{G}_i:=\hat{G}_i-\langle \hat{G}_i \rangle_n$ on the state $\ket{n(x)}$ can be written as
\begin{equation}
\Delta \hat{G}_i\ket{n(x)}=\left(\rmi\hbar\partial_{i}-\langle\hat{G}_i\rangle_{n}\right)|n\rangle=i\hbar\left(\partial_{i}+iA_{i}^{(n)}\right)|n\rangle,
\end{equation}
which resembles the structure of the covariant derivative $D_{i}^{(n)}=\partial_{i}+iA_{i}^{(n)}$ with connection $A_{i}^{(n)}$.

In the following example we shall see that  Eqs.~(\ref{QG:QIM}) and (\ref{QG:A}) yield the expected results.

\subsection*{Example: Generalized harmonic oscillator with a linear term}

For this example we consider the quantum generalized harmonic oscillator with a linear term described by the Hamiltonian operator~(\ref{ghoWq:quantumH}). The notation used here is the same as in Example B of Sec.~\ref{Examples}. The starting point is to promote the corresponding classical functions $G_i$ given by (\ref{ghoWq:Gi}) to the quantum operators: 
\begin{equation}\label{ghoWq2:Goperator}
\hat{G}_i(\hat{q},\hat{p};x)\!=\!\frac{1}{2}f_i(x)\! \left( \hat{q}\hat{p}+\hat{p}\hat{q} \right) + g_i(x) \hat{q}^2 + h_i(x) \left( \hat{p} + \frac{Y}{Z}\hat{q}  \right)  ,
\end{equation}
where $f_i(x)$, $g_i(x)$, and $h_i(x)$ are given by Eqs.~(\ref{ghoWq:f}), (\ref{ghoWq:g}), and (\ref{ghoWq:h}), respectively. Notice that by construction~(\ref{ghoWq2:Goperator}) is Hermitian. Then, using the eigenfunctions~(\ref{ghoWq:wavefunctgen}) and the properties of the Hermite functions~(\ref{Hermite}), we compute the quantum metric tensor~(\ref{QG:QIM}) with the operators~(\ref{ghoWq2:Goperator}), obtaining
\begin{eqnarray}\label{ghoWq2:QIM}
&&g^{(n)}_{ij}(x) = \frac{1+n+n^2}{2}\left(f_i(x) f_j(x) + \frac{Z^2}{\omega^2} l_i(x) l_j(x) \right) \nonumber\\
&& +\frac{\left(n+\frac{1}{2}\right)\omega}{\hbar Z} \left(\frac{4W^2 Z^4}{\omega^6} l_i(x) l_j(x)  + m_i(x) m_j(x) \right)\!, \ \ \ \ \
\end{eqnarray}
where $l_i(x):=g_i(x)-\frac{Y}{Z} f_i(x)$ and $m_i(x):=h_i(x)-\frac{W Z}{\omega^2} f_i(x)$. It can be verified, by explicit calculation, that Eq.~(\ref{ghoWq2:QIM}) leads directly to Eq.~(\ref{ghoWq:QIM}), which corroborates the validity of Eq.~(\ref{QG:QIM}). Finally, we apply Eq.~(\ref{QG:A}) to the operators (\ref{ghoWq2:Goperator}). The result is
\begin{equation}
A^{(n)}_i(x) = \left( g_i(x)-\frac{Y}{Z} f_i(x) \right) \left[\left(n+\frac{1}{2}\right)\frac{Z}{\omega} + \frac{W^2 Z^2}{\hbar \omega^4}\right],
\end{equation}
which leads to Eq.~(\ref{ghoWq:Bconnetion}) and, hence, verifies the validity of Eq.~(\ref{QG:A}).

\section{Conclusion}\label{sec:Conclusions}

In this paper, we have introduced the metric (\ref{gclas:metric}) for classical integrable systems and shown through examples that it corresponds to the classical counterpart of the quantum metric tensor~(\ref{QIM}). The classical metric is defined on the parameter space and provides a measure of the distance between nearby points in phase space, which is induced by the adiabatic evolution of the classical system. We investigate the main features of this classical metric. In particular, we show that this metric is gauge invariant in the parameter space in the sense that  it remains unchanged when we perform  the canonical transformation~(\ref{gclas:gauge}), meaning that this classical metric is independent of the ``zero'' point from which we measure the angle variables. Most importantly, we find for the considered examples that this metric agrees with the quantum metric tensor in rank and the functional dependence on the parameters. This allowed us to establish the exact relation between both metrics for the generalized harmonic oscillator and the generalized harmonic oscillator with a linear term, provided the Bohr-Sommerfeld quantization rule for action variable. For the nontrivial example of the quartic anharmonic oscillator, these metrics were calculated by using perturbation theories, and hence we find an approximate relation between them. We use the generating function (\ref{classical:G2}) of translations  in the parameter space as the fundamental object to build the aforementioned classical metric and demonstrated that Hannay's curvature could also be expressed in terms of it; thereby providing a unified treatment for both geometric structures. Finally, we extend the use of this classical generating function to the quantum case and obtain alternative expressions for the quantum metric tensor and Berry's connection, which are verified for the case of the quantum generalized harmonic oscillator with a linear term. 

We would like to close by pointing out some remarks. First, it would be interesting to address the possible existence of the classical analog of the non-Abelian quantum metric tensor proposed in Ref.~\cite{YuQuan2010}, and the generalization of the metric~(\ref{gclas:metric}) for the case of classical systems with chaotic dynamics along the lines of Ref.~\cite{Robbins631}. In particular, those authors first study the quantum case, where they resort to time dependence to find an expression for Berry's curvature. Then they introduce a semiclassical approximation and get an expression for the classical curvature, where instead of employing action-angle variables, they perform the integration restricted to a particular energy shell. This classical curvature then reduces to Hannay's curvature when the system is integrable. In this spirit, as both the quantum and classical cases are still tractable, we may as well generalize  our proposed classical metric. Nevertheless, for a nonintegrable Hamiltonian that slightly differs from an integrable Hamiltonian, the classical metric~(\ref{gclas:metric}) might shed some light on the chaotic behavior through the application of the canonical perturbation theory. A further generalization of Eq.~(\ref{gclas:metric}) is one wherein the classical metric is invariant under a more general gauge transformation where the shift $\lambda$ in Eq.~(\ref{gclas:gauge}) also depends on the angle variables, this motivated by the work of Ref.~\cite{Alvarez-Jimenez2016}. Another interesting and useful future consideration is how to generalize the metric~(\ref{gclas:metric}) for a classical field theory. 

Apart from possible generalizations, the metric~(\ref{gclas:metric}), being the classical analog of the quantum metric tensor, may help to provide more insight into the investigation of quantum phase transitions. Furthermore, the  metric~(\ref{gclas:metric}) may be relevant in the context of shortcuts to adiabatic processes in classical integrable systems, which consist of the use of a control Hamiltonian $K_c(\varphi,I;x)$ that turns out to be $K_c= G_i(\varphi,I;x) \dot{x}^i$ and achieves a constant action  variable $I$ with arbitrarily fast changing parameters $x$~\cite{Deng2013}. In this line of thought, it may be noted that in Ref.~\cite{Bravetti2017} is proposed a metric analogous to Eq.~(\ref{gclas:metric}) that emerges in the study of the thermodynamic cost of shortcuts to adiabaticity and defines a distance between the initial and final statistical states of the classical system.

\acknowledgments

We thank Juan Carlos Del Valle for his valuable help in computing the wave function and the spectrum for the quantum quartic anharmonic oscillator. This work was partially supported by DGAPA-PAPIIT Grants No. IN103716 and No. IN103919, CONACyT project 237503. Daniel Guti\'errez-Ruiz is supported with a CONACyT Ph.D. scholarship (No. 332577). Diego Gonzalez is supported with a DGAPA-UNAM postdoctoral fellowship.

\appendix*

\section{Quantum quartic anharmonic oscillator}

The Hamiltonian operator of the system is
\begin{eqnarray}
\hat{H}=\frac{\hat{p}^{2}}{2m}+\frac{k}{2}\hat{q}^{2}+ \frac{\lambda}{4!}\hat{q}^{4}.
\end{eqnarray}
Here we consider a perturbative treatment in the parameter $\lambda$ since there is not an exact solution to the resulting Schr\"odinger equation. Furthermore, we restrict ourselves to obtain the ground-state wave function and its energy up to third order in $\lambda$. To accomplish this task, we will use the  method proposed in Ref.~\cite{Turbiner1984}, which is suitable to find corrections to large powers of $\lambda$.

The adimensional form of the eigenvalue problem considered 
is
\begin{equation}
\left(-\frac{d^{2}}{dQ^{2}}+Q^{2}+\Lambda Q^{4}\right)\psi(Q)=\epsilon\psi(Q),
\end{equation}
where we defined the following quantities
\begin{equation}
\Lambda:=\frac{\hbar\lambda}{12m^{2}\omega_{0}^{3}}, \qquad  Q:=\sqrt{\frac{m\omega_{0}}{\hbar}}q,\qquad \epsilon:=\frac{2E}{\hbar\omega_{0}},
\end{equation}
with $\omega_{0}=(k/m)^{1/2}$. After applying the method, the resulting nonnormalized ground-state wave function $\psi_0(Q)$, to third order in $\Lambda^{3}$, is
\begin{eqnarray}
\psi_0(Q;x)&&=e^{\frac{-Q^{2}}{2}}\bigg[1-\frac{\Lambda Q^{2}}{8}\left(Q^{2}+3\right)+\frac{\Lambda^{2}Q^{2}}{384}\left(3Q^{6}\right.\nonumber\\
&&\left.+26Q^{4}+93Q^{2}+252\right)-\frac{\Lambda^{3}Q^{2}}{3072}\left(Q^{10}+17Q^{8}\right.\nonumber\\
&&\left.+141Q^{6}+813Q^{4}+2916Q^{2}+7992\right)\bigg],
\end{eqnarray}
and the corresponding energy is
\begin{equation}
\epsilon=1+\frac{3}{4}\Lambda-\frac{21}{16}\Lambda^{2}+\frac{333}{64}\Lambda^{3}.
\end{equation}

Returning to our original variables and normalizing, we arrive at
\begin{eqnarray}\label{quartic:wave}
&&\psi_0(q;x)=e^{-\frac{m\omega_{0}}{2\hbar}q^{2}}\bigg[\sqrt[4]{\frac{m\omega_{0}}{\pi\hbar}}-\frac{\lambda P_{1}(q;x)}{384\sqrt[4]{\pi m^{7}\omega_{0}^{11}\hbar^{5}}}\nonumber \\ &&+\frac{\lambda^{2}P_{2}(q;x)}{884736\sqrt[4]{\pi m^{15}\omega_{0}^{23}\hbar^{9}}}
\!-\! \frac{\lambda^{3}P_{3}(q;x)}{339738624\sqrt[4]{\pi m^{23}\omega_{0}^{35}\hbar^{13}}}\bigg], \ \ \ \ \
\end{eqnarray}
and
\begin{equation}
E_{0}=\frac{\hbar\omega_{0}}{2}+\frac{\hbar^{2}\lambda}{32m^{2}\omega_{0}^{2}}-\frac{7\hbar^{3}\lambda^{2}}{1536m^{4}\omega_{0}^{5}}+\frac{37\hbar^{4}\lambda^{3}}{24576m^{6}\omega_{0}^{8}},\label{eq:enerquart}
\end{equation}
where
\begin{eqnarray}
P_{1}(q;x)&=&4m^{2}\omega_{0}^{2}q^{4}+12\hbar m\omega_{0}q^{2}-9\hbar^{2},\\
P_{2}(q;x)&=&48m^{4}\omega_{0}^{4}q^{8}+416\hbar m^{3}\omega_{0}^{3}q^{6}+1272\hbar^{2}m^{2}\omega_{0}^{2}q^{4}\nonumber\\
&&+3384\hbar^{3}m\omega_{0}q^{2}-4677\hbar^{4},\\
P_{3}(q;x)&=&64m^{6}\omega_{0}^{6}q^{12}+1088\hbar m^{5}\omega_{0}^{5}q^{10}+8592\hbar^{2}m^{4}\omega_{0}^{4}q^{8}\nonumber\\
&&+48288\hbar^{3}m^{3}\omega_{0}^{3}q^{6}+154524\hbar^{4}m^{2}\omega_{0}^{2}q^{4}\nonumber\\
&&+419076\hbar^{5}m\omega_{0}q^{2}-729153\hbar^{6}.
\end{eqnarray}
Plugging Eq.~(\ref{quartic:wave}) into Eq.~(\ref{QIM}) and keeping terms up to second order in $\lambda$ (due to derivatives with respect to $\lambda$ are present), we arrive at the components of the quantum metric tensor given by Eq.~(\ref{quartic:QIM}).

\bibliography{references}

\end{document}